# Tropical rainforest bird community structure in relation to altitude, tree species composition, and null models in the Western Ghats, India

T. R. SHANKAR RAMAN, N. V. JOSHI, AND R. SUKUMAR*

*Centre for Ecological Sciences, Indian Institute of Science, Bangalore 560012, India*

**\*For correspondence:**

R. SUKUMAR

Centre for Ecological Sciences

Indian Institute of Science

Bangalore 560012

INDIA

Tel.: +91 80 2360 0382

Fax.: + 91 80 2360 2280

E-mail: rsuku@ces.iisc.ernet.in

# Abstract

Studies of species distributions on elevational gradients are essential to understand principles of community organisation as well as to conserve species in montane regions. This study examined the patterns of species richness, abundance, composition, range sizes, and distribution of rainforest birds at 14 sites along an elevational gradient (500-1400 m) in the Kalakad-Mundanthurai Tiger Reserve (KMTR) of the Western Ghats, India. In contrast to theoretical expectation, resident bird species richness did not change significantly with elevation although the species composition changed substantially (<10% similarity) between the lowest and highest elevation sites. Constancy in species richness was possibly due to relative constancy in productivity and lack of elevational trends in vegetation structure. Elevational range size of birds, expected to increase with elevation according to Rapoport's rule, was found to show a contrasting inverse U-shaped pattern because species with narrow elevational distributions, including endemics, occurred at both ends of the gradient (below 800 m and above 1,200 m). Bird species composition also did not vary randomly along the gradient as assessed using a hierarchy of null models of community assembly, from completely unconstrained models to ones with species richness and range-size distribution restrictions. Instead, bird community composition was significantly correlated with elevation and tree species composition of sites, indicating the influence of deterministic factors on bird community structure. Conservation of low- and high-elevation areas and maintenance of tree species composition against habitat alteration are important for bird conservation in the southern Western Ghats rainforests.

# Introduction

Since the pioneering work of Terborgh (1971, 1977, 1985, Terborgh and Weske 1975), research on altitudinal distributional patterns of birds has tended to focus on three major aspects—variation with altitude in richness, range sizes, and species turnover rates—and on one region with relatively well-documented information on bird altitudinal distributions, the Neotropics (Rahbek 1997, Patterson *et al.* 1998, Stotz 1998, Blake and Loiselle 2000). Most studies have shown a general pattern of decreasing species richness with increasing altitude, believed to mirror the well-recognised latitudinal gradient in species richness (MacArthur 1972, Patterson, Pacheco and Solari 1996, Patterson *et al.* 1998, Bachman *et al.* 2004). Monotonic decline in species richness with altitude may also occur when altitudinal range sizes of species increase with altitude, as suggested by Stevens (1992) in an extension of Rapoport's rule, a matter of intense debate (Rapoport 1982, Gaston, Blackburn and Spicer 1998, Taylor and Gaines 1999).

Other studies have and suggested that species richness may show a non-monotonic, hump-shaped pattern with diversity peaking at mid-altitudes (Janzen 1973, Rahbek 1997, Stotz 1998), as in diversity-productivity gradients (Rosenzweig and Abramsky 1993). This pattern may become apparent when data are corrected for sampling effort for local species richness (Terborgh 1977, Patterson *et al.* 1996), or area for regional species richness (Rahbek 1997, Bachman *et al.* 2004). Models that evaluate consequences of geometric constraints (hard upper and lower limits) on the altitudinal ranges of species also predict mid-altitude peaks called the 'mid-domain effect' (Rahbek 1997, Colwell and Lees 2000, Veech 2000).

In contrast to species richness, community compositional change (turnover rates) at the regional level may increase with altitude (Rahbek 1997), or show peaks and troughs corresponding to transition zones between lowland and montane avifaunas (Patterson *et al.*

1996, Stotz 1998). Little is known of factors influencing bird community composition at local levels in tropical rainforest. Along successional gradients, habitat structure and tree species composition influence bird community structure, wherein structurally and floristically similar sites have more similar bird communities (Raman *et al.* 1998). Along altitudinal gradients, past studies of local bird community structure have largely ignored the relative influence of deterministic differences between sites in habitat attributes versus the effects of chance, null models, or altitude *per se*.

The present study explores variation in local bird community structure, species composition, and turnover rates along an altitudinal gradient in a tropical rainforest of the Western Ghats of India. Although recognised as a global biodiversity hotspot (Myers *et al.* 2000), and an endemic bird area (Stattersfield *et al.* 1998), this region is peculiar in being relatively depauperate in its rainforest bird community due to historical isolation from larger tracts of rainforest in northeast India and southeast Asia (Daniels *et al.* 1992). Here, we explore how bird community attributes vary with altitude in tropical rainforest. Specifically, we ask: does species richness show a monotonic decline with altitude or a hump-shaped pattern due to the 'mid-domain effect'? Is there support for Rapoport's rule? Does altitudinal distance and degree of change in habitat influence and correlate with change in bird community composition or are turnover rates estimated accurately by null models that simulate non-equilibrium dynamics? The results are used to assess the relative influence of deterministic *versus* non-equilibrium factors on tropical rainforest bird community structure.

# Materials and methods

**Study area**

The 1600 km long chain of hills called the Western Ghats runs along the west coast of the Indian peninsula from 8° N to 21° N. Moist forests, including tropical wet evergreen forest, are found largely south of 16° N (Pascal 1988), and contain a higher diversity of endemic plant and animal taxa (Ali and Ripley 1983, Daniels 1992, Kumar *et al.* 1999). The fieldwork was carried out in the Kalakad-Mundanthurai Tiger Reserve (KMTR, 895 km², 8°25′ to 8°53′ N, 77°10′ to 77°35′ E) in the Agasthyamalai region at the southern end of the Western Ghats. KMTR ranges between 50 m and 1700 m above sea level (asl) with rainforest occurring chiefly above 500 m. This reserve along with adjoining areas has one of the largest remaining contiguous tracts (over 400 km²) of relatively undisturbed tropical rainforest in the Western Ghats (Ramesh *et al.* 1997). The rainforest vegetation in KMTR has been classified mid-elevation tropical wet evergreen rainforest of the *Cullenia exarillata-Mesua ferrea-Palaquium ellipticum* type (Pascal 1988, Ganesh *et al.* 1996, Parthasarathy 2001). Within the rainforest, the mean monthly daytime temperature ranges between 19° C in January and 24° C in April–May (at Sengaltheri, 1,040 m, range 15°–31° C). The total rainfall was 2283 mm in 1998 and 2230 mm in 1999 at Sengaltheri. There are three seasons: (a) dry season (February to May), (b) southwest or summer monsoon (June to September), and (c) northeast or winter monsoon (October to January). KMTR receives over half its annual precipitation during the northeast monsoon.

Although 278 bird species have been recorded in and around KMTR, only 84 species occur in rainforests including 12 endemics and 13 winter migrants (Raman 2001). A majority of species breed between late January and May (Ali and Ripley 1983, TRSR personal observations). The sites chosen for intensive bird community sampling were located at Kannikatti (740 m, 8°37′ N and 77°16′ E), Sengaltheri (1040 m, 8°31′ N and 77°26′ E), and Kakachi (1220 m, 8°33′ N and 77°24′ E) in KMTR. These areas contain rainforests between 500 m and 1400 m altitude, contiguous with rainforests elsewhere in the Reserve. Fourteen sites located along this altitudinal gradient in relatively undisturbed rainforest were selected randomly, within logistical constraints, around three forest camps (four accessed from Kannikatti, six from Sengaltheri, and four from Kakachi) and located on topographic maps using landmarks and a global positioning system receiver (GPS).

**Bird surveys**

The fixed-radius point count method was used to survey bird populations in each site in a relatively uniform and efficient manner during the main breeding season (Verner 1985, Ralph *et al.* 1995). Point count surveys (5 min duration) were carried out during the first three hours after sunrise (see Raman 2003 for details of field technique). Densities estimated by the fixed radius approach were used as they were highly correlated to variable-radius point count estimates across species (Raman 2003).

Each sampled site represented an area of around 12 – 15 ha lying at the designated altitude and was surrounded by contiguous rainforests on at least three sides. Within each site a 600 – 700 m long transect or narrow animal trail was marked at 25 m intervals for the point count surveys. At each site, 25 point count surveys (yielding 167 – 247 detections and an estimated 334 – 597 individual birds per site) were carried out, excluding the Kodayar site where only 18 counts were carried out. Sampling of all sites around Sengaltheri and Kakachi was carried out mostly between February and May 1998, during the peak breeding season when resident and migrant birds were present. For reasons of logistics, the four sites around Kannikatti were sampled only in March 1999.

**Vegetation sampling**

In each of the 14 sites, densities of trees greater than 30 cm girth at breast height (GBH at 1.3 m) were estimated using the point-centred quarter method (PCQ, Krebs 1989). Thirteen PCQ plots, with successive plots spaced 50 m apart, were measured in each site, giving a sample of 52 trees per site. Comparison with two completely enumerated 1 ha plots (>1800 stems each) in Sengaltheri showed that the density estimates from this PCQ sample was <2% different from the more laborious enumeration. All trees were identified to species, or in a few cases to genus (Gamble and Fischer 1915-1935, Pascal and Ramesh 1997). Distance from plot centre to the middle of the bole and GBH were recorded for each tree. At each of the 13 PCQ plots, 2 m radius plots were laid to enumerate shrubs. In addition, the number of cane plants and the presence or absence of bamboo within 5 m radius was recorded.

Altitude, canopy measures, and leaf litter variables were measured at 25 points, evenly spaced 25 m apart, in each site. Canopy height was measured using a rangefinder and percentage canopy cover using a spherical densiometer. Vertical stratification was assessed at these 25 points by noting the presence or absence of foliage was noted in the following height intervals (in metres): 0–1, 1–2, 2–4, 4–8, 8–16, 16–24, 24–32, and > 32, directly above and in a 0.5 m radius around each point (Raman *et al.* 1998). Leaf litter depth on the forest floor was measured using a calibrated wooden probe. As ground vegetation and litter were disturbed along trails, the samples were taken 10 m away from trails into the rainforest interior.

**Data analysis**

The cumulative list of bird species recorded in each site was used as a basic measure of bird species richness. However, because only 18 point counts were sampled at the Kannikatti site, we obtained standardised estimates of bird community parameters for all sites for 17 sampled points. Using the program EstimateS (Colwell 1997), 100 permutations (17 samples without replacement) were performed to estimate the following parameters and their standard deviations: (1) bird species richness, measured as the cumulative number of species; (2) bird abundance, estimated as the number of individual birds/ha; (3) Shannon-Weiner diversity, calculated as $-\Sigma(p_i \ln p_i)$, where $p_i$ is the proportion of the i[th] species and the summation is across all species in the pooled samples; (4) rarefaction richness, standardising for sampling effort using Coleman rarefaction curves (Colwell 1997); (5) bootstrap species richness, an incidence-based estimator denoted as $S_{boot} = S_{obs} + \Sigma (1 - p_k)^m$, where $p_k$ is the proportion of $m$ samples with species $k$ and the summation is across all $S_{obs}$ species in the pooled samples; and (6) Chao 1 richness, an abundance-based estimator of species richness (see Colwell 1997 for the formula and correct variance estimator of this index).

Smaller scale point richness and abundance estimates were also obtained by averaging across replicate point count surveys in each site. The average number of species, detections, and individuals per point and their standard errors were estimated for all species combined and for resident species alone. For each site, tree density and basal area were calculated using the PCQ method (Krebs 1989). Average values across replicate sampling points in each site were calculated for other vegetation and site variables: shrub, cane, and cardamom densities, leaf litter depth, canopy height and cover, and altitude. Vertical stratification was measured as the average number of strata with foliage across the 25 points sampled in each site. The coefficient of variation of this index indexed horizontal heterogeneity (following Raman *et al.* 1998). The total number of tree species recorded in the PCQ plots was recorded as a measure of tree species richness. Relationships between bird community and vegetation variables were assessed using Kendall rank correlations (Siegel and Castellan 1988).

Altitudinal range size of each species was estimated as the difference between the maximum and minimum altitude at which the species was recorded in the point count surveys. The altitudinal range midpoint and a weighted range midpoint, estimated using the abundance of each species in each site, were calculated for each species. Similarities in bird community, foliage profile, and tree species composition between sites were computed using the Morisita index that is least sensitive to sample size effects (Wolda 1981, Raman *et al.* 1998). The matrix of pairwise dissimilarities (100−Morisita index in %) in bird community composition was related to corresponding matrices of similarity/distance in tree species composition, foliage profile, altitude, and geographic distance between sites using Mantel tests (Manly 1994). Geographic distance between sites was measured as the straight-line distance between sampling sites. Statistical significance of Mantel tests was assessed through 10000 random permutations. To examine the independent effects of different variables, partial Mantel tests, derived from the Kendall tau approach, were used (Hemelrijk 1990). The tree and bird community similarity matrices were also used to ordinate study sites using multidimensional scaling (Manly 1994).

We used correlation and regression techniques to assess statistical significance of altitudinal trends. Linear and non-linear (chiefly quadratic) regressions were applied to describe spatial trend data. We also used non-parametric Kendall (*tau*) rank-order correlation coefficients to assess statistical significance (Siegel and Castellan 1988).

**Null model analyses**

The mid-domain effect (Colwell and Lees 2000) was assessed using the null model of Veech (2000). Species' altitudinal range widths were retained but located randomly on the altitudinal gradient between 500 and 1400 m (divided into 18 zones of 50 m width) in 1000 simulations to obtain an expected species richness curve along this gradient. The observed pattern of species richness was compared to the simulated curve to obtain a mean displacement, $D$, where $D = (\sum d_j)/18$, where $d_j$ is the absolute difference in species richness between the two curves being compared for each of the 18 zones (Veech 2000). A further 1000 simulations of the data were used to determine the distribution of $D$ values in each run and to assess statistical significance.

Null models were also used to assess the likelihood of obtaining, under the null model assumptions, estimates equivalent to two measures of the average similarity between sites in bird community composition. These measures were: (i) average similarity between all possible pairs of sites ($N$ = 91 comparisons), and (ii) average similarity between successive sites along the altitudinal gradient ($N$ = 13 comparisons). The Jaccard index (Krebs 1989) was used to measure similarity in bird community composition. With the presence-absence matrix of species across sites, five null models were evaluated:

**1.** Unconstrained model: matrix total constant, row and column totals varied;

**2.** Saturation model: fixed column totals (species richness at each site);

**3.** Distribution model: fixed row totals (number of sites each species occurred in);

**4.** Doubly constrained model: fixed row and column totals by a procedure equivalent to the Knight's tour algorithm (Sanderson 2000);

**5.** Range-contiguity model: this model simulated, in addition to the distribution constraint, the contiguity of ranges of species along the gradient. Simulations showed that species occurrences were distributed more contiguously along the gradient than expected by random placement ($P < 0.001$). Using the observed occurrences in the species-by-site matrix, the relative probability of occurrence in any one of the 12 intermediate sites (excluding the 2 'edge' sites—the lowest and highest site) was found to be enhanced by a factor of 9.0 if the species also occurred in sites immediately above and below and by a factor of 4.5 if it occurred only in one adjacent site either above or below. These enhancement factors were in relation to the probability of occurrence in a site when the species was absent from both adjacent sites on the gradient. For the edge sites, occurrence was enhanced by a factor of 3.6 if the species occurred in the adjacent site in relation to occurrence when it was absent in the adjacent site. These were then used to distribute occurrences in simulations to produce probabilistic contiguous distributions.

These models attempt to simulate the effects of chance and biological constraints on the assembly of species communities (Farnsworth and Ellison 1996, Stone *et al.* 1996, Sanderson 2000). In contrast to the unconstrained model, the saturation, distribution, and doubly-constrained models implement constraints on the number of species that can fill each site and simulate the effects of interspecific differences in distribution and occurrence probabilities. Our range-contiguity model incorporates, in addition, constraints on species' altitudinal distributions and contiguity of ranges. We retained in the contiguity model the number of intermediate absences or 'holes' in range but randomised their location. Even with these constraints, we checked and established that over 5000 different matrices existed in simulations. For each of the above models, we simulated 5000 randomised matrices and the averages of the obtained distribution of similarity values (between all possible pairs of sites and adjacent sites) were statistically compared to the observed average similarity values.

## Results

**Bird species richness, diversity, and abundance**

Across the 14 sites, we obtained over 2900 detections comprising approximately 6600 individuals of birds belonging to 67 species. Of these, nine species were latitudinal (winter) migrants and the remaining were resident species. Thirteen species were detected only once and a further seven species were seen less than five times overall. Total bird species richness (cumulative list in each site) was significantly negatively correlated to altitude of the site (Kendall rank-order correlation, $\tau = -0.52$, $n = 14$, $P = 0.01$, Table 1). Resident bird species richness appeared to decline with altitude, but this trend was not significant statistically ($P = 0.08$). The standardised estimates for 17 point counts obtained over 100 runs (sampling with replacement) of species richness, Shannon-Weiner diversity, Coleman rarefaction richness, bootstrap richness, and the Chao1 richness were all significantly negatively correlated to altitude ($P < 0.05$, Table 1). Most of the decrease occurred between 500 m and 900 m altitude, after which there was little change in species richness. Bird abundance showed a converse positive association with altitude that was marginally significant ($\tau = 0.39$, $n = 14$, $P = 0.055$, Table 1). The per point estimates of total and resident bird species richness, detections, and abundance showed significant variation across the 14 sites (one-way ANOVA, $F_{13,329} > 5.58$, $P < 0.004$). Whereas the number of detections per point showed weak negative correlations with altitude ($\tau = -0.41$, $P = 0.04$), the trends for total and resident species richness and detections with altitude were non-significant. Bird abundance per point increased with altitude, weakly and non-significantly in the case of total abundance ($\tau = 0.38$, $P = 0.06$) and significantly in the case of abundance of resident species ($\tau = 0.52$, $P = 0.009$).

**Vegetation and birds: correlations**

Vegetation variables were not significantly correlated with altitude (Kendall correlations, $P > 0.05$), except for bamboo culm density ($\tau = 0.52$, $n = 14$, $P = 0.01$) mainly due to the occurrence of bamboo at only two higher altitude sites. The six standardised estimates of bird species richness, Shannon-Weiner bird diversity, and the per-point total and resident bird species richness were all significantly negatively correlated to one of the vegetation variables: leaf-litter depth ($\tau < -0.45$, $n = 14$, $P < 0.03$ in all cases). These variables were also not correlated to any of the other vegetation variables. Bird abundance (standardised and per-point) and the total number of bird detections per point were not correlated to any vegetation variable. The average number of resident bird detections per point was negatively correlated to leaf litter depth ($\tau = -0.47$, $P = 0.037$).

**Range distributions, sizes, and species richness**

The range size and distribution pattern across the 58 resident bird species varied from species occurring from one to all 14 sites spanning the entire gradient. Twenty-three species (39.7%) were either restricted to lower altitudes or showed declining trends with altitude in population density (Kendall correlations, $P < 0.10$, Appendix). Sixteen species (27.6%) displayed a trend of increase in population density with altitude ($P < 0.10$) or were restricted largely to higher altitude sites (Appendix). Only two species of woodpeckers, Common Flameback and White-bellied Woodpecker, appeared to be restricted to mid-altitudes. The spatial distributions of the remaining species were widespread and relatively uniform (14 species) or could not be determined due to low sample sizes (3 species).

Across the gradient, the number of resident bird species whose ranges intercepted each 50 m altitude zone was nearly constant at around 35 species (Fig. 1a). This was significantly different from the Veech (2000) null model that predicted a mid-domain peak (Fig. 1a, $d = 6.51$, $P < 0.001$). The relationship between altitudinal range size of a species and the weighted midpoint of its altitudinal range was also non-linear and quadratic (Fig. 1b, fitted curve: $y = -0.0046\ x^2 + 8.8633\ x - 3496.8$, $R^2 = 0.72$). Similarly, in contrast to the expectation under Rapoport's rule, the mean altitudinal range of species in each site showed a quadratic relationship with the altitude of the site (Fig. 1c, fitted curve: $y = -0.0005\ x^2 + 8.8839\ x + 216.99$, $R^2 = 0.78$).

**Bird species turnover**

Species turnover was considerable along the altitudinal gradient. Non-metric multidimensional scaling ordination using the Morisita dissimilarity matrix of the data on tree and bird species composition showed similar patterns (stress = 0.070 and 0.034, Fig. 2a–b, respectively). Sites below 900 m tended to cluster together as did sites above 1,100 m, while intermediate sites were relatively dissimilar to the others (with exceptions shown in tree species composition by site D at 843 m and site M at 1,341 m).

Partial Mantel tests showed that dissimilarity in bird community composition between sites was independently positively correlated to both altitudinal distance between sites ($T = 0.60$, $P < 0.0001$, Fig. 3a) and dissimilarity in tree species composition ($T = 0.22$, $P = 0.013$, Fig. 3b). Correlations with both these variables were significant even when the effects of geographic distance were controlled for ($T = 0.69$ and $0.51$, $P < 0.0003$). In contrast, geographic distance between sites had no effect on bird species composition, when the effects of altitudinal distance were controlled for ($T = 0.08$, $P = 0.12$), indicating the primary importance of altitude. We found no significant effect of foliage profile dissimilarity on bird community composition.

Species turnover between sites (average of 91 all-possible pairs) was $0.529 \pm 0.139$ SD by the Jaccard similarity index and $0.608 \pm 0.251$ SD by the Morisita similarity index. This was lower than average similarity in bird community composition between adjacent sites (Jaccard index: $0.693 \pm 0.107$ SD, Morisita index: $0.860 \pm 0.103$ SD). The pattern of turnover was not monotonic or increasing with altitude, and remained relatively low but for higher turnover (lower similarity) around 1,000 m (Fig. 4).

**Null models of species turnover**

We assessed whether the five null models of bird species turnover could accurately estimate the observed Jaccard similarity between all possible pairs of sites ($0.529 \pm 0.139$ SD) and between adjacent pairs of sites ($0.693 \pm 0.107$ SD). The unconstrained model and saturation models gave similar results, producing similarity estimates that were significantly lower than the observed ($P < 0.0002$, Fig. 5). The estimated average similarities between all possible pairs of sites by the range-contiguity model and the double-constraint model were close to that actually observed (0.516 *vs.* 0.529) but was, nevertheless, significantly lower ($P < 0.0002$). The double-constraint model also performed poorly in estimating average similarity between adjacent sites (0.517 *vs.* 0.693). The range-contiguity model estimated similarities between adjacent sites that were higher and much closer to the observed value (0.622 *vs.* 0.693). However, the model estimates were significantly lower than the observed values for double-constraint and range contiguity models ($P < 0.0002$, Fig. 5).

## Discussion

### Species richness

Species richness of resident rainforest birds varied little despite substantial change in community composition across the altitudinal gradient, in contrast to expectation from the four main models hitherto proposed (Rahbek 1997). The general model, propounding a monotonous decline in species richness with altitude in parallel with an assumed decline in productivity, finds weak support, if any, when all bird species including latitudinal migrants are considered. The pattern of range sizes observed in this study results in local species richness patterns that depart significantly from expectations under Rapoport's rule, hump-shaped relationships with productivity (Rosenzweig and Abramsky 1993), or mid-altitude peaks arising from geometric constraints (the mid-domain effect, Colwell and Lees 2000).

Instead, the observed pattern of bird species richness is a spatial analogue of the recent theoretical model proposed by Brown *et al.* (2001), to explain the regulation of local species richness over time in changing environments. This model suggests that species composition may vary substantially over time, but species richness, as an emergent property of ecosystems, is often regulated within narrow limits. Five conditions are necessary and sufficient for this to occur (Brown *et al.* 2001): (i) productivity, or resource availability, remains relatively constant, (ii) other abiotic or biotic factors vary, causing turnover in species composition, (iii) a regional species pool provides a source of colonists to local sites that are open systems with respect to species colonisation and extinction, (iv) the species pool contains species capable of utilizing the entire range of resources and showing compensatory shifts in abundance, and (v) the division rule governing apportioning of resources across species results in similar ranked species-abundance distributions.

There is partial support for these criteria for rainforest birds in KMTR. Although productivity could not be directly measured, the relative lack of clinal change in vegetation variables across the altitudinal gradient suggests that habitat structure and resource availability did not vary substantially. This is partly because a narrower altitudinal gradient was sampled in this study (500–1,400 m) compared to studies from the Neotropics (Terborgh 1971, 1977, Rahbek 1997). Over a wider range bird diversity may declines with altitude and the corresponding reduction in forest structural complexity as in the Peruvian Andes (Terborgh 1977). In the present study, bird species richness variables were uncorrelated to any vegetation variables except for a negative, possibly spurious, correlation with leaf litter depth. The second condition of Brown *et al.* (2001), species turnover due to changes in abiotic and biotic variables, finds clear support in the correlated variation in bird community composition in relation to altitude and tree species composition. The existence of a species pool that is nearly twice as large (with at least 58 species) as in any one local sampling site (around 30 species), adduces support for the third criterion. Finally, the occurrence of a number of species that occurred across the entire altitudinal gradient and the similar overall bird abundance and ranked species-abundance distributions (data not presented here) suggest support for the final criteria. Further support is, however, required regarding compensatory shifts in abundance of bird species along the altitudinal gradient, possibly in relation to occurrence of other bird species (Terborgh and Weske 1975).

**Species range sizes and distribution**

Range sizes of rainforest birds at KMTR do not conform to the expectations under Rapoport's rule, which may operate only under specific ecological conditions including competition and the rescue effect (Taylor and Gaines 1999). The failure of Rapoport's rule and the mid-domain effect is consequential to the peculiar distribution pattern of rainforest birds in KMTR. A number of species, particularly those endemic to the Western Ghats, were restricted in distribution to lower (e.g. Malabar Grey Hornbill, White-bellied Blue Flycatcher) or higher (Black-and-Orange Flycatcher, White-bellied Shortwing) altitudes (Appendix, Ali and Ripley 1983). Daniels (1992) has noted that a significant proportion of endemics, among birds and angiosperms in the Western Ghats, is restricted to the higher hills (> 1000 m). In other words, bird distributions were mainly of three types, restricted-range species of low and high altitudes, and widely distributed species, with no mid-altitude species with small ranges. The low altitude species included many that also use moist deciduous forest habitats (e.g., Malabar Grey Hornbill, Malabar Parakeet, Grey-headed Bulbul), whereas high-altitude restricted species are largely confined to wet evergreen rainforest (e.g., Black-and-Orange Flycatcher, White-bellied Shortwing, Nilgiri Flycatcher, Ali and Ripley 1983). This pattern may be a consequence of historical factors that influenced the prevalence and distribution of rainforest over geological time scales.

**Species turnover**

A spectrum of opinion exists on the factors influencing the composition or assembly of species communities at particular sites and their variation over space and time. At one extreme, equilibrial models suggest that local communities are integrated, repeatable units whose composition is strictly regulated and predictable as a result of deterministic factors such as the varying environmental tolerances of species and competition (Clements 1916,

MacArthur 1972, Pandolfi 1996, Terborgh *et al.* 1996, Pitman *et al.* 2001). In contrast, non-equilibrium approaches note that community composition varies substantially over space and time, apparently due to stochastic or historical effects of colonization and extinction (Gleason 1926, Whittaker 1970, Strong *et al.* 1984, Hubbell and Foster 1986, Brown *et al.* 2001). A fundamental distinction between non-equilibrium and equilibrium models is that under the former, community composition may be expected to 'drift' or vary continuously through space and time, whereas the latter predicts that spatially or temporally independent sites with similar environmental conditions would have similar communities (Hubbell and Foster 1986, Terborgh *et al.* 1996). Using a space-for-time substitution approach, biological influences and deterministic structure have been demonstrated for communities of tropical rainforest trees (Terborgh *et al.* 1996, Pitman *et al.* 2001) and birds (Terborgh and Weske 1975, Raman *et al.* 1998).

The significant effect on bird community composition of altitude and tree species composition, and the lack of influence of geographic distance, suggests the inapplicability of non-equilibrium models of randomly-varying distributions-abundance patterns with spatial dependence (Terborgh *et al.* 1996). Thus, high-altitude sites in Sengaltheri clustered with high-altitude sites farther away (Neterikal trail, Kakachi, and Kodayar) than to virtually adjacent sites at lower altitudes. Bird community composition may be constrained by the altitude-specific environment including temperature, irradiance, and other biological factors known to vary with altitude in tropical rainforests (Richards 1996). In addition, tree species composition appear to be a key determinant of rainforest bird community composition in this study as in other studies from south-western India (Raman and Sukumar 2002), and north-eastern India (Raman *et al.* 1998).

The pattern of species turnover with altitude indicated the occurrence of distinctive community composition at low-altitudes (< 900 m) and high altitudes (> 1,100 m) separated by a transitional zone of high turnover. This paralleled the observed range size distributions and placements of rainforest birds in KMTR. The occurrence of the transition zone at around 1,000–1,200 m altitude may be due to environmental changes related to the formation of cloud- and mist-cover during the monsoon months in the southern Western Ghats as in other tropical montane rainforest regions (Richards 1996).

**Null models of community assembly**

In large-scale community studies, models with few or no constraints that attempt to parsimoniously simulate the effects of pure chance, almost invariably fail to explain community structure (Farnsworth and Ellison 1996, Terborgh *et al.* 1996, Pitman *et al.* 2001). In this study, we evaluated a hierarchy of null models with constraints that attempt to inject varying degrees of biological realism. The results clearly indicate that non-equilibrium models that do not incorporate essential biological constraints fail to predict or reproduce the observed pattern of similarities in species composition between sites. Even with constraints on site richness, species' range widths, and contiguity, the results suggest that assuming that species are distributed independently of each other can result in community similarities close to, but departing significantly from, those actually observed. Thus, Whittaker's (1970) models of local community structure as a consequence of independent overlapping species distributions along ecological gradients, is only partly supported. The data suggests the possibility that some species have significantly higher overlap or avoidance due to biological determinants such as joint distributions up to ecotones or exclusive distributions due to competitive interactions (Terborgh 1971, 1985). Field research in the Western Ghats on competition and on avian distributional ecology along with null model tests (e.g., Hofer *et al.* 1999, 2000), may shed more light on these aspects.

## Acknowledgements

This research was funded by the Ministry of Environment and Forests, India, and John D. and Catherine T. MacArthur Foundation, USA. We thank the Tamil Nadu Forest Department for research permissions and V. K. Melkani, for his interest and support in the field. N. M. Ishwar, Divya Mudappa, and K. Vasudevan were helpful and inspiring colleagues and we thank them and Ravi Chellam for unstintingly sharing the field station resources. Divya helped enormously in fieldwork, data collection, and discussions. We are grateful to P. Jeganathan for assistance with vegetation data collection and to U. Hofer and M. B. Krishna for discussions. M. Jeyapandian, A. Silamban, Sashikumar, and many others are thanked for their assistance in the field.

# References

Ali, S. and S.D. Ripley (1983): Handbook of the birds of India and Pakistan. Compact edition. Oxford University Press, Delhi.

Bachman, S., W.J. Baker, N. Brummitt, J. Dransfield and J. Moat (2004): Altitudinal gradients, area and tropical island diversity: an example from the palms of New Guinea. *Ecography 27*: 299-310.

Blake, J.G. and B.A. Loiselle (2000): Diversity of birds along an altitudinal gradient in the Cordillera Central, Costa Rica. *Auk 117*: 663-686.

Brown, J.H., S.K.M. Ernest, J.M. Parody and J.P. Haskell (2001): Regulation of diversity: maintenance of species richness in changing environments. *Oecologia 126*: 321−332.

Clements, F.E. (1916): Plant succession: An analysis of the development of vegetation. Publication No. 242, Carnegie Institute, Washington DC.

Colwell, R.K. (1997): EstimateS: Statistical estimation of species richness and shared species from samples. Version 6.01b. URL: http://viceroy.eeb.uconn.edu/EstimateS.

Colwell, R.K.and D.C. Lees (2000): The mid-domain effect: geometric constraints on the geography of species richness. *Trends in Ecology and Evolution 15*: 70−76.

Daniels, R.J.R. (1992): Geographical distribution patterns of amphibians in the Western Ghats, India. *Journal of Biogeography 19*: 521−529.

Daniels, R.J.R., N.V. Joshi and M. Gadgil (1992): On the relationship between bird and woody plant species diversity in the Uttara Kannada District of south India. *Proceedings of the National Academy of Sciences (USA) 89*: 5311–5315.

Farnsworth, E.J. and A.M. Ellison (1996). Scale-dependent spatial and temporal variability in biogeography of mangrove root epibiont communities. *Ecological Monographs 66*: 45−66.

Gamble, J.S. and C.E.C. Fischer (1915-1935). Flora of the Presidency of Madras. Parts I to XI. Secretary of State for India, London.

Ganesh, T., R. Ganesan, M.S. Devy, P. Davidar and K.S. Bawa (1996). Assessment of plant biodiversity at a mid-altitude evergreen forest of Kalakad-Mundanthurai Tiger Reserve, Western Ghats, India. *Current Science 71*: 379–392.

Gaston, K.J., T.M. Blackburn and J.I. Spicer (1998). Rapoport's rule: time for an epitaph? *Trends in Ecology and Evolution 13*: 70–74.

Gleason, H.A. (1926). The individualistic concept of plant association. *Bulletin of the Torrey Botanical Club 53*: 7–26.

Grimmett, R., C. Inskipp and T. Inskipp (1998). Birds of the Indian subcontinent. Oxford University Press, Delhi.

Hemelrijk, C.K. (1990). Models of, and tests for, reciprocity, unidirectionality and other social interaction patterns at a group level. *Animal Behaviour 39*: 1013–1029.

Hofer, U., L.-F. Bersier and D. Borcard (1999). Spatial organization of a herpetofauna on an altitudinal gradient revealed by null model tests. *Ecology 80*: 976–988.

Hofer, U., L.-F. Bersier and D. Borcard (2000). Ecotones and gradient as determinants of herpetofaunal community structure in the primary forest of Mount Kupe, Cameroon. *Journal of Tropical Ecology 16*: 517–533.

Hubbell, S.P. and R.B. Foster (1986). Biology, chance, and the history and structure of tropical rain forest tree communities. Pp 314-329. *In*: Community ecology (Eds.: Diamond, J.M. & T.J. Case), Harper and Row, New York.

Inskipp, T., N. Lindsey and W. Duckworth (1996). An annotated checklist of the birds of the Oriental region. Oriental Bird Club, Sandy, Bedfordshire.

Janzen, D. (1973). Sweep samples of tropical foliage insects: effects of seasons, vegetation types, altitude, time of day, and insularity. *Ecology 54*: 687–708.

Krebs, C.J. (1989). Ecological methodology. Harper and Row, New York.

Kumar, A., W.R. Konstant and R.A. Mittermeier (1999). Western Ghats and Sri Lanka. Pp. 354-365. *In*: Hotspots: Earth's biologically richest and most endangered terrestrial ecoregions (Eds.: Mittermeier, R.A., N. Myers & C.G. Mittermeier). CEMEX Conservation International, Mexico.

MacArthur, R.H. (1972). Geographical ecology: patterns in the distribution of species. Harper and Row, New York.

Manly, B.F.J. (1994). Multivariate statistical methods: a primer, 2nd edn. Chapman and Hall, London.

Myers, N., R.A. Mittermeier, C.G. Mittermeier, G.A.B . da Fonseca and J. Kent (2000). Biodiversity hotspots for conservation priorities. *Nature 403*: 853–858.

Pandolfi, J.M. (1996). Limited membership in Pleistocene reef coral assemblages from the Huon Peninsula, Papua New Guinea: constancy during global change. *Paleobiology 22*: 152-176.

Parthasarathy, N. (2001). Changes in forest composition and structure in three sites of tropical evergreen forest around Sengaltheri, Western Ghats. *Current Science 80*: 389-393.

Pascal, J.P. (1988). Wet evergreen forests of the Western Ghats of India: ecology, structure, floristic composition and succession. Institut Français de Pondichéry, Pondicherry, India.

Pascal, J.P. and B.R. Ramesh (1997). A field key to the trees and lianas of the evergreen forests of the Western Ghats (India). Institut Français de Pondichéry, Pondicherry, India.

Patterson, B.D., V. Pacheco and S. Solari (1996). Distribution of bats along an altitudinal gradient in the Andes of south-eastern Peru. *Journal of Zoology 240*: 637-658.

Patterson, B.D., D.F. Stotz, S. Solari and J.W. Fitzpatrick (1998). Contrasting patterns of altitudinal zonation for birds and mammals in the Andes of southeastern Peru. *Journal of Biogeography 25*: 693-607.

Pitman, N.C.A., J.W. Terborgh, M.R.P. Silman, V. Núñez, D.A. Neill, C.E. Cerón, W.A. Palacios and M. Aulestia (2001). Dominance and distribution of tree species in upper Amazonian terra firme forests. *Ecology 82*: 2101-2117.

Rahbek, C. (1997). The relationship among area, altitude, and regional species richness in Neotropical birds. *American Naturalist 149*: 875–902.

Ralph, C.J., J.R. Sauer and S. Droege (Eds.) (1995). Monitoring bird populations by point counts. USDA, Forest Service, Pacific Southwest Research Station, Albany, California.

Rapoport, E.H. (1982). Areography: geographical strategies of species. Pergamon Press, Oxford.

Raman, T.R.S. (2001). Community ecology and conservation of tropical rainforest birds in the southern Western Ghats, India. Ph. D. thesis, Indian Institute of Science, Bangalore.

Raman, T.R.S. (2003). Assessment of census techniques for interspecific comparisons of tropical rainforest bird densities: a field evaluation in the Western Ghats, India.*Ibis 145*: 9-21.

Raman, T.R.S. and Sukumar, R. (2002). Responses of tropical rainforest birds to abandoned plantations, edges and logged forest in the Western Ghats, India. *Animal Conservation 5*: 201-216.

Raman, T.R.S., G.S. Rawat and A.J.T. Johnsingh (1998). Recovery of tropical rainforest avifauna in relation to vegetation succession following shifting cultivation in Mizoram, north-east India. *Journal of Applied Ecology 35*: 214–231.

Ramesh, B.R., S. Menon and K.S. Bawa (1997). A vegetation-based approach to biodiversity gap analysis in the Agasthyamalai Region, Western Ghats, India. *Ambio 26*: 529-536.

Richards, P.W. (1996). The tropical rain forest: an ecological study, 2nd edn. Cambridge Univ. Press, Cambridge.

Rosenzweig, M.L. and Z. Abramsky (1993). How are diversity and productivity related? Pp. 52-65. *In*: Species diversity in ecological communities: historical and geographical perspectives (Eds: Ricklefs, R. & D. Schluter). University of Chicago Press, Chicago.

Sanderson, J.G. (2000). Testing ecological patterns. *American Scientist 88*: 332–339.

Siegel, S. and N.J. Castellan Jr. (1988). Nonparametric statistics for the behavioral sciences. McGraw-Hill, New York.

Stattersfield, A.J., M.J. Crosby, A.J. Long and D.C. Wege (1998). Endemic bird areas of the world: priorities for biodiversity conservation. Birdlife International, Cambridge.

Stevens, G.C. (1992). The altitudinal gradient in altitudinal range: an extension of Rapoport's rule to altitude. *American Naturalist 140*: 893–911.

Stone, L., T. Dayan and D. Simberloff (1996). Community-wide assembly patterns unmasked: the importance of species' differing geographical ranges. *American Naturalist 148*: 997-1015.

Stotz, D.F. (1998). Endemism and species turnover with altitude in montane avifaunas in the neotropics: implications for conservation. *In*: Conservation in a changing world (Eds: Mace, G., A. Balmford & R. Joshua). Pp. 161–186. Cambridge University Press, Cambridge.

Strong, D.R., D. Simberloff, L.G. Abele and A.B. Thistel (eds). (1984). Ecological communities: conceptual issues and the evidence. Princeton University Press, Princeton.

Taylor, P.H. and S.D. Gaines (1999). Can Rapoport's rule be rescued? Modeling causes of the latitudinal gradient in species richness. *Ecology 80*: 2474–2482.

Terborgh, J. (1971). Distribution on environmental gradients: theory and a preliminary interpretation of distributional patterns in the avifauna of the Cordillera Vilcabamba, Peru. *Ecology 52*: 23−40.

Terborgh, J. (1977). Bird species diversity on an Andean altitudinal gradient. *Ecology 58*: 1007−1019.

Terborgh, J. (1985). The role of ecotones in the distribution of Andean birds. *Ecology 66*: 1237−1246.

Terborgh, J. and J.S. Weske (1975). The role of competition in the distribution of Andean birds. *Ecology 56*: 562-576.

Terborgh, J., R.B. Foster and V.P. Nuñez (1996). Tropical tree communities: a test of the nonequilibrium hypothesis. *Ecology 77*: 561−567.

Veech, J.A. (2000). A null model for detecting nonrandom patterns of species richness along spatial gradients. *Ecology 81*: 1143−1149.

Verner, J. (1985). Assessment of counting techniques. *Current Ornithology 2*: 247-302.

Whittaker, R.H. (1970). Communities and ecosystems. Macmillan, New York.

Wolda, H. (1981). Similarity indices, sample size and diversity. *Oecologia 50*: 296-302.

TABLE 1. Bird community parameters along an altitudinal gradient in the rainforests of Kalakad-Mundanthurai Tiger Reserve, India, showing Kendall rank-order correlations with altitude.

| Transect Code | A | B | C | D | E | F | G | H | I | J | K | L | M | N | Correlation with altitude | | |
|---|---|---|---|---|---|---|---|---|---|---|---|---|---|---|---|---|---|
| Mean altitude (m) | 558 | 646 | 751 | 843 | 900 | 980 | 1092 | 1228 | 1256 | 1259 | 1265 | 1287 | 1341 | 1359 | τ | Z | P |
| **Cumulative species richness** | | | | | | | | | | | | | | | | | |
| Total species | 40 | 36 | 36 | 34 | 30 | 29 | 31 | 29 | 32 | 31 | 30 | 25 | 28 | 32 | -0.520 | -2.59 | 0.0096 |
| Resident species | 35 | 30 | 30 | 29 | 26 | 25 | 27 | 26 | 30 | 27 | 28 | 25 | 25 | 30 | -0.352 | -1.75 | 0.0798 |
| **Standardised estimates (EstimateS)** | | | | | | | | | | | | | | | | | |
| Bird species richness | 36.8 | 33.0 | 33.0 | 31.1 | 27.4 | 25.9 | 28.2 | 27.0 | 28.9 | 27.5 | 26.3 | 21.9 | 27.6 | 29.97 | -0.407 | -2.03 | 0.0428 |
| SD | 1.60 | 1.46 | 1.60 | 1.34 | 1.55 | 1.99 | 1.38 | 1.31 | 1.37 | 1.83 | 1.90 | 1.72 | 0.60 | 1.15 | | | |
| Bird abundance (individuals/ha) | 22.6 | 24.3 | 21.7 | 22.3 | 23.0 | 17.0 | 22.1 | 29.7 | 24.8 | 25.5 | 20.7 | 25.2 | 25.8 | 30.5 | 0.385 | 1.92 | 0.0554 |
| SD | 0.81 | 1.10 | 1.20 | 1.18 | 1.64 | 1.51 | 1.61 | 1.50 | 1.63 | 1.58 | 1.48 | 2.15 | 0.68 | 1.93 | | | |
| Shannon-Weiner diversity | 3.22 | 3.03 | 3.06 | 3.02 | 2.73 | 2.74 | 2.82 | 2.69 | 2.79 | 2.63 | 2.48 | 2.07 | 2.53 | 2.8 | -0.626 | -3.12 | 0.0018 |
| SD | 0.04 | 0.04 | 0.05 | 0.04 | 0.08 | 0.08 | 0.05 | 0.07 | 0.06 | 0.07 | 0.11 | 0.09 | 0.03 | 0.05 | | | |
| Coleman rarefaction richness | 37.7 | 34.1 | 34.2 | 32.3 | 28.5 | 27.8 | 29.8 | 28.1 | 31.4 | 29.1 | 27.9 | 23.4 | 27.8 | 31.3 | -0.538 | -2.68 | 0.0073 |
| SD | 1.29 | 1.18 | 1.16 | 1.14 | 1.06 | 0.99 | 0.96 | 0.84 | 0.72 | 1.23 | 1.25 | 1.10 | 0.41 | 0.76 | | | |
| Bootstrap richness | 41.1 | 37.1 | 37.3 | 34.3 | 30.8 | 29.6 | 31.5 | 29.8 | 32.3 | 31.2 | 30.4 | 25.2 | 30.5 | 33.1 | -0.429 | -2.14 | 0.0328 |
| SD | 2.11 | 1.92 | 2.15 | 1.79 | 1.98 | 2.45 | 1.80 | 1.71 | 1.86 | 2.42 | 2.46 | 2.19 | 0.76 | 1.42 | | | |
| Chao1 richness | 40.6 | 33.9 | 35.1 | 32.1 | 28.3 | 26.2 | 28.5 | 27.2 | 29.0 | 28.3 | 28.3 | 23.6 | 28.3 | 30.2 | -0.420 | -2.09 | 0.0365 |
| SD | 6.31 | 2.06 | 4.38 | 2.08 | 2.20 | 0.86 | 1.19 | 0.19 | 0.46 | 1.60 | 3.70 | 4.38 | 2.08 | 1.00 | | | |

LEGENDS TO FIGURES

**Fig. 1.** Bird species richness and range size relationships in rainforests of KMTR; (a) upper panel: evaluation of the Veech (2000) null model of the mid-domain effect—line shows best fit curve to null model estimates against observed curve (dotted line); (b) middle panel: altitudinal range size of 58 resident bird species in relation to the weighted mid-point of their altitudinal range; and (c) lower panel: mean altitudinal range of species at each of the 14 sampled sites in relation to altitude of the sites.

**Fig. 2.** Ordination of study sites on the basis of similarity in (a) tree species composition (upper panel) and (b) bird community composition (lower panel) using multidimensional scaling. Sites are named A to N in increasing order of altitude.

**Fig. 3.** Bird community dissimilarity between sites (91 all-possible pairs) in relation to corresponding between-site (a) altitudinal distance (upper panel) and (b) tree species dissimilarity (lower panel).

**Fig. 4.** Turnover in rainforest bird community composition as a function of altitude in KMTR. In the upper panel (a) similarity between each site and the immediately higher site is plotted against altitude of the lower site. The lower panel (b) uses data from 10 roughly equally-spaced sites along the gradient and plots rate of change in similarity between a site and the sites immediately lower and higher to it against altitude of the central site. The rate of change of similarity was calculated as $½[(d_1/w_1)+(d_2/w_2)]$, where $d_1$ and $w_1$ are the Morisita index dissimilarity and altitudinal distance between a site and the lower site, and $d_2$ and $w_2$ the corresponding values for the higher site.

**Fig. 5.** Between-site similarities (average Jaccard index) in rainforest bird community composition estimated from the five non-equilibrium null models compared with observed values in the data set. Vertical lines are 1 SD.

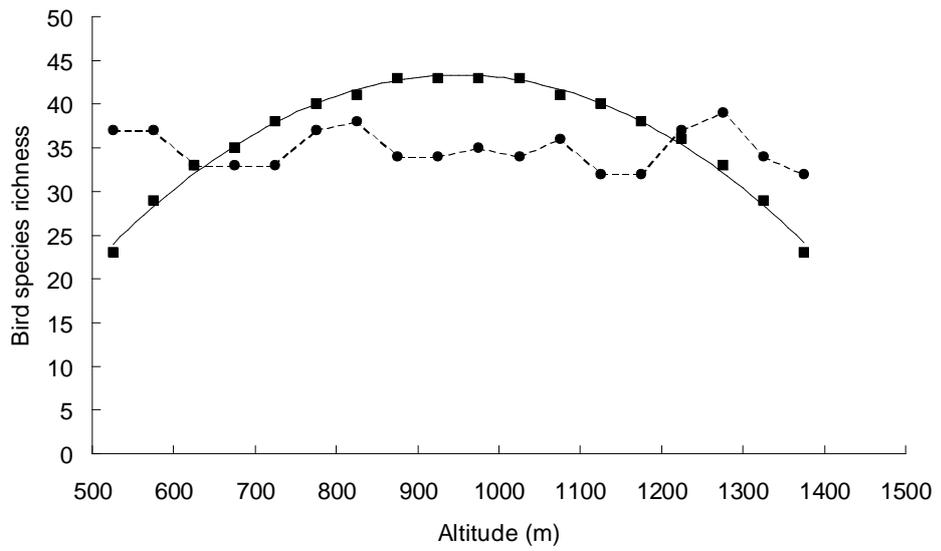
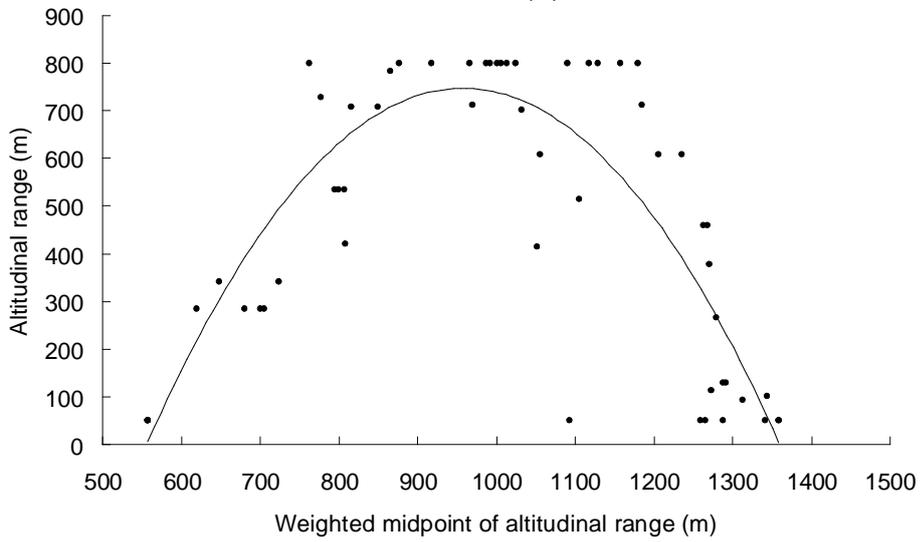
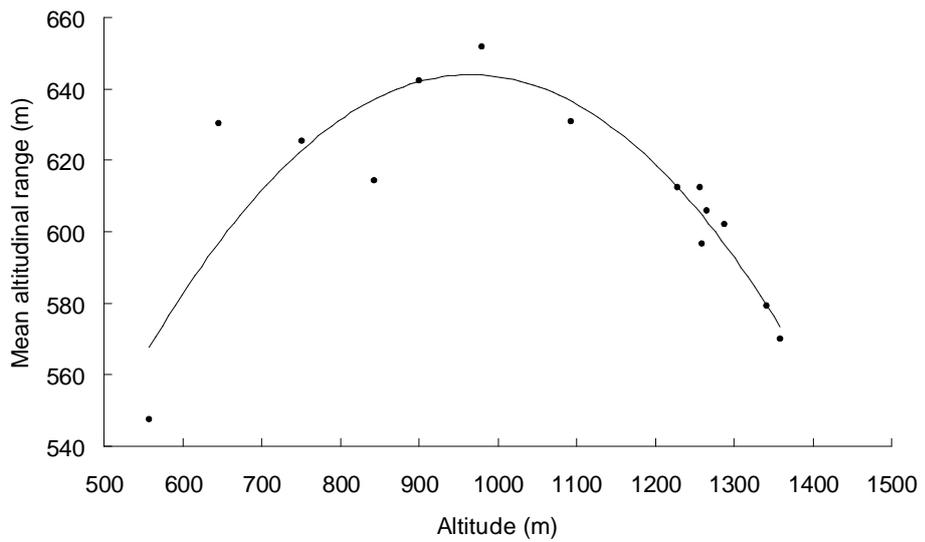

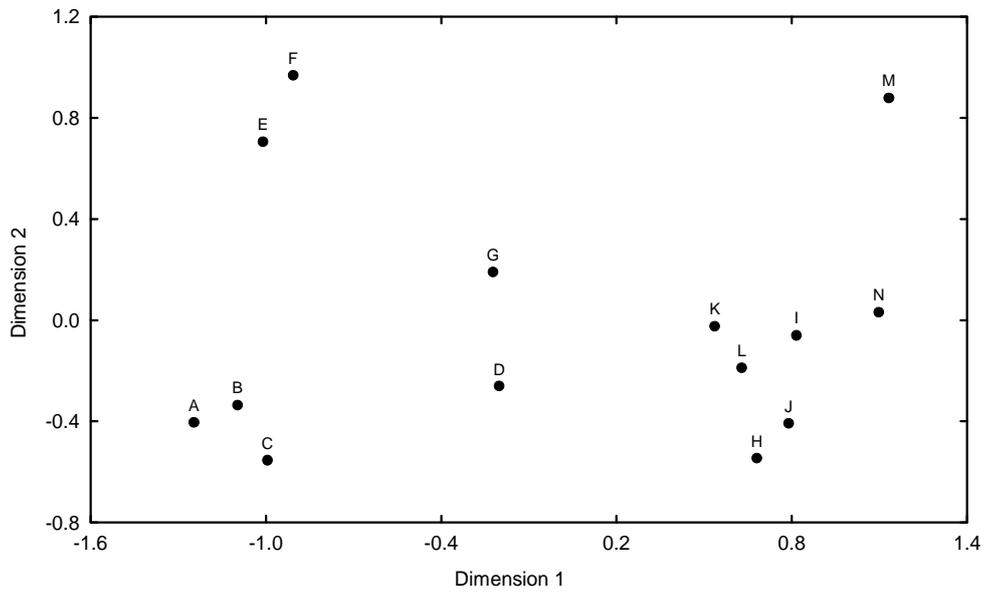

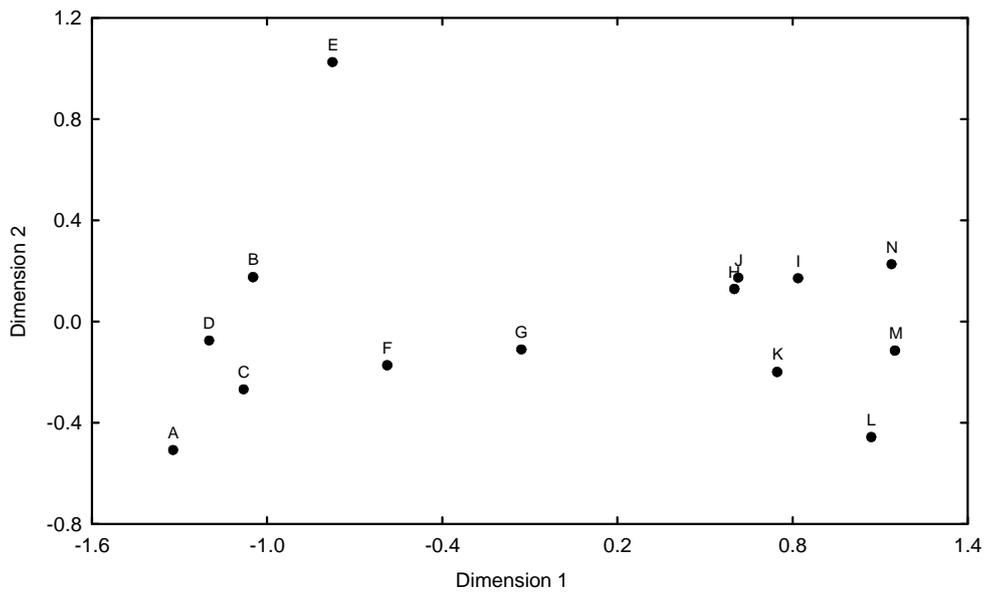

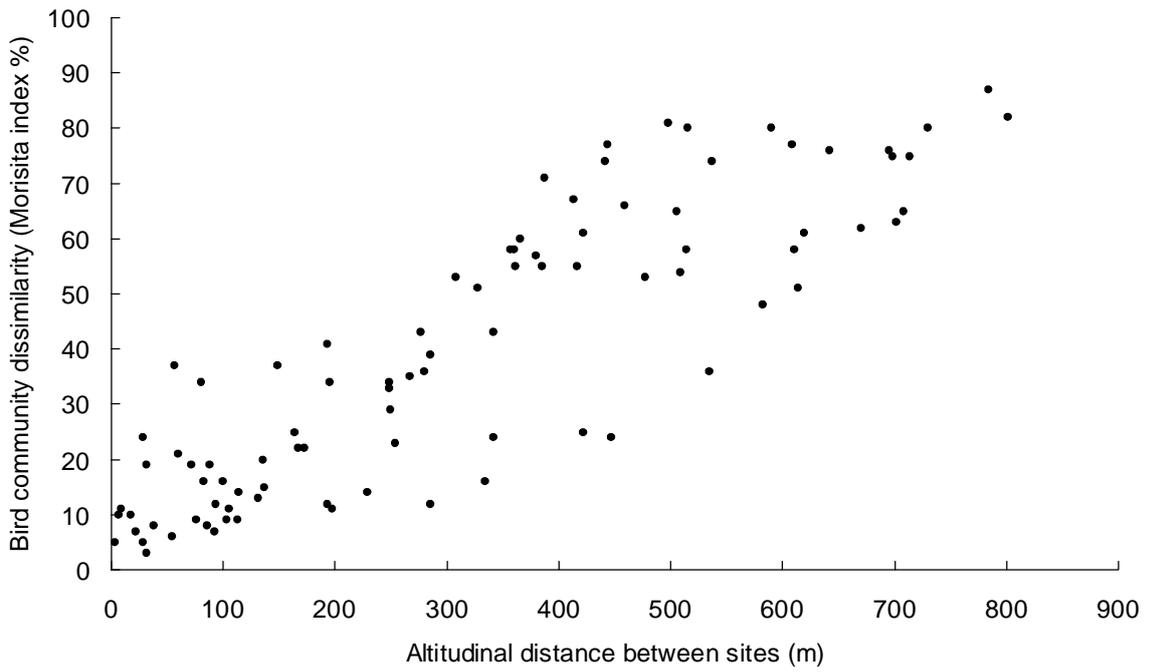

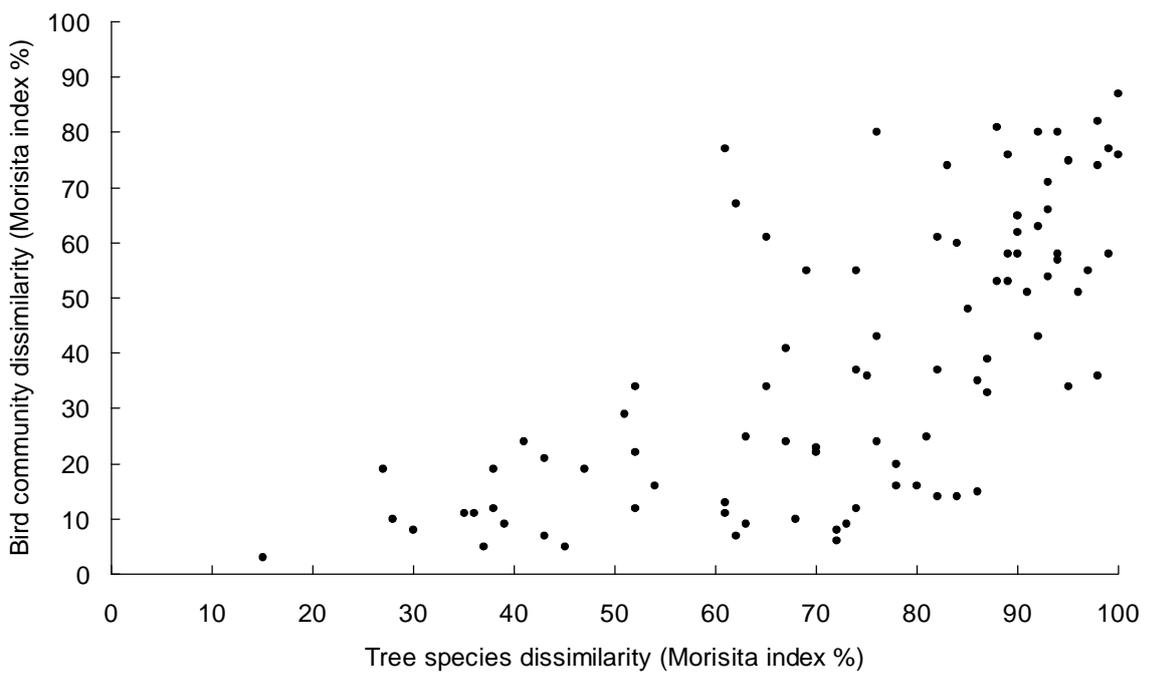

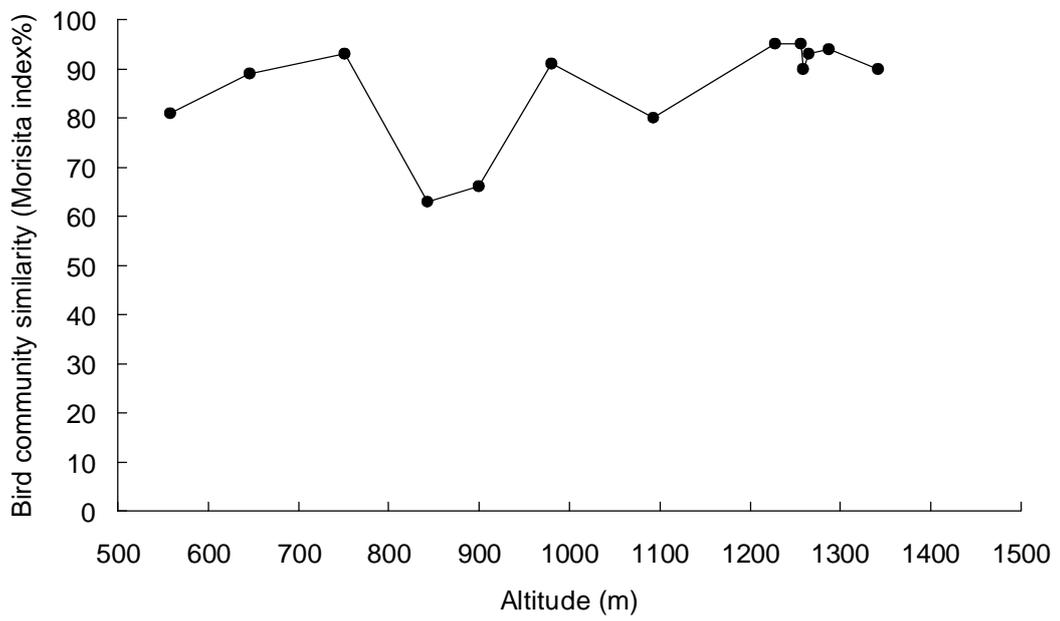

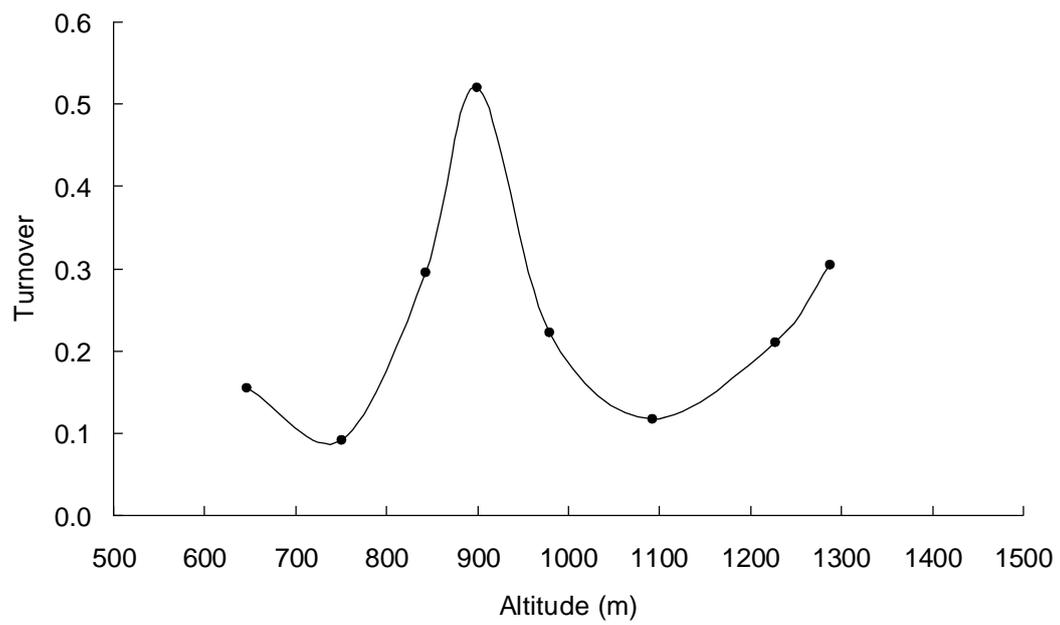

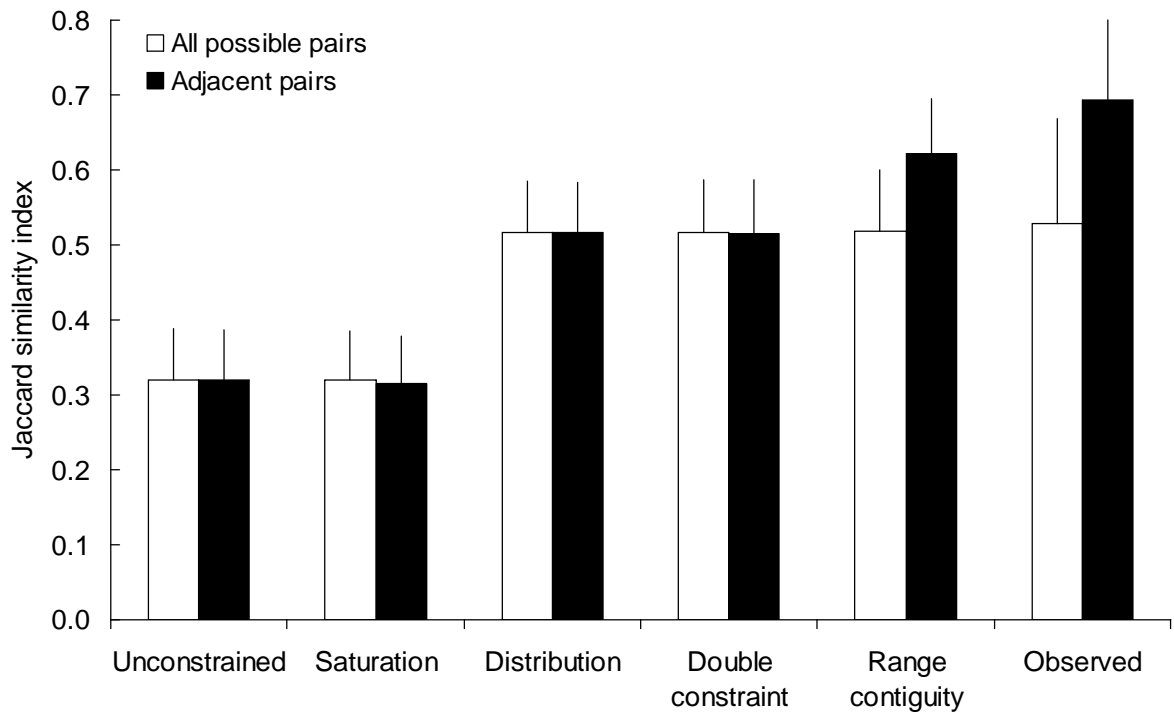

## Appendix

Range distribution and abundance (individuals/ha) of resident rainforest bird species in KMTR. Species are arranged increasing order of elevational mid-point. Rank refers to increasing correlation (Kendall Tau) between site elevation and abundance across sites A to N (see Table 1). Species names follow Inskipp, Lindsey and Duckworth (1996) and Grimmett, Inskipp and Inskipp (1998).

| Common name | A | B | C | D | E | F | G | H | I | J | K | L | M | N | Tau | P | Rank |
|---|---|---|---|---|---|---|---|---|---|---|---|---|---|---|---|---|---|
| Large Woodshrike | 0.1 | | | | | | | | | | | | | | -0.38 | 0.060 | 17 |
| Grey-headed Bulbul* | 0.1 | | | | | | | | | | | | | | -0.38 | 0.060 | 18 |
| Loten's Sunbird | 0.1 | | | | | | | | | | | | | | -0.38 | 0.060 | 19 |
| Plum-headed Parakeet | 0.1 | | | | | | | | | | | | | | -0.38 | 0.060 | 20 |
| Rufous Babbler* | 0.2 | | | | | | | | | | | | | | -0.38 | 0.060 | 21 |
| Crimson-fronted Barbet | 0.2 | 0.1 | | 0.1 | | | | | | | | | | | -0.58 | 0.004 | 10 |
| Hill Myna | 0.5 | 0.3 | 0.1 | 0.6 | | | | | | | | | | | -0.62 | 0.002 | 6 |
| Malabar Grey Hornbill* | 0.6 | 0.2 | 0.2 | 0.4 | | | | | | | | | | | -0.62 | 0.002 | 7 |
| Bronzed Drongo | 0.1 | | | 0.1 | | | | | | | | | | | -0.43 | 0.033 | 15 |
| Black-crested Bulbul | 0.7 | 0.1 | | 0.3 | 0.5 | | | | | | | | | | -0.56 | 0.006 | 12 |
| Gold-fronted Leafbird | 0.3 | 0.1 | 0.1 | | 0.1 | | | | | | | | | | -0.64 | 0.001 | 4 |
| White-bellied Treepie | 0.6 | 0.6 | 0.5 | 1.8 | 0.8 | 0.7 | | | | | | | | | -0.59 | 0.004 | 9 |
| Asian Fairy Bluebird | 1.0 | 1.1 | 0.5 | 0.9 | 0.6 | 0.8 | 0.8 | | | | | | | | -0.69 | 0.001 | 3 |
| Malabar Parakeet* | 0.8 | 0.5 | 0.3 | 0.8 | 0.2 | 0.2 | 0.6 | | | | | | | | -0.71 | 0.000 | 1 |
| Black-naped Monarch | 0.3 | 0.3 | 0.4 | 0.4 | 0.9 | | 0.1 | | | | | | | | -0.58 | 0.004 | 11 |
| Rufous Woodpecker | 0.1 | | | | 0.1 | 0.2 | 0.1 | 0.1 | 0.1 | 0.1 | | | | | -0.21 | 0.291 | 28 |
| Besra | 0.2 | 0.1 | 0.1 | | | 0.1 | | | | | 0.1 | | | | -0.45 | 0.026 | 14 |
| Greater Racket-tailed Drongo | 2.0 | 1.1 | 1.1 | 1.2 | 1.1 | 0.4 | 0.5 | 0.4 | 0.2 | 0.7 | 0.5 | | | | -0.70 | 0.000 | 2 |

| Common name | A | B | C | D | E | F | G | H | I | J | K | L | M | N | Tau | P | Rank |
|---|---|---|---|---|---|---|---|---|---|---|---|---|---|---|---|---|---|
| Vernal Hanging Parrot | 0.8 | 0.4 | | 0.1 | | 0.4 | 0.3 | | 0.1 | | 0.1 | 0.1 | | | -0.43 | 0.034 | 16 |
| White-bellied Blue Flycatcher* | 1.1 | 0.7 | 0.7 | 1.1 | 0.8 | 1.1 | 0.3 | | 0.1 | | 0.4 | 0.2 | 0.1 | | -0.53 | 0.009 | 13 |
| Greater Flameback | 0.7 | 0.5 | 0.7 | 0.3 | 0.1 | | 0.2 | 0.5 | 0.3 | 0.1 | 0.4 | 0.4 | 0.3 | 0.3 | -0.23 | 0.254 | 27 |
| Mountain Imperial Pigeon | 0.8 | 1.5 | 1.5 | 1.3 | 0.3 | 0.1 | 0.2 | | | | | | | 0.2 | -0.63 | 0.002 | 5 |
| Grey Junglefowl | 0.5 | 0.2 | 0.7 | 0.6 | 0.4 | 0.1 | 0.1 | 0.2 | 0.3 | 0.1 | | | | 0.1 | -0.62 | 0.002 | 8 |
| Crimson-backed Sunbird* | 3.1 | 2.8 | 3.2 | 3.3 | 3.0 | 2.7 | 2.9 | 3.2 | 1.2 | 2.8 | 2.9 | 3.1 | 0.4 | 1.2 | -0.37 | 0.065 | 22 |
| Malabar Trogon | 0.3 | 0.5 | 0.3 | | 0.2 | 0.2 | | 0.3 | 0.2 | 0.1 | 0.2 | | 0.1 | 0.2 | -0.35 | 0.080 | 23 |
| Puff-throated Babbler | 0.4 | 0.5 | 0.8 | 0.9 | 0.3 | 0.5 | 0.3 | 0.4 | 0.9 | 0.2 | 0.5 | 0.2 | 0.4 | 0.3 | -0.28 | 0.162 | 24 |
| White-cheeked Barbet | 0.5 | 1.0 | 0.3 | 0.7 | 0.1 | 0.4 | 1.2 | | | | 0.1 | 0.1 | 0.3 | 0.3 | -0.27 | 0.171 | 25 |
| Yellow-browed Bulbul | 1.3 | 2.1 | 2.3 | 1.8 | 2.3 | 2.4 | 2.8 | 1.0 | 0.6 | 0.6 | 1.1 | 1.6 | 1.1 | 1.1 | -0.26 | 0.203 | 26 |
| Scarlet Minivet | 0.9 | 0.4 | 0.7 | 0.1 | 0.4 | 0.8 | 1.0 | 0.2 | 0.4 | | 0.2 | 0.3 | 0.5 | 0.7 | -0.12 | 0.543 | 29 |
| Velvet-fronted Nuthatch | 0.5 | 0.8 | 0.2 | 0.3 | 0.5 | | | 0.3 | 0.3 | 0.1 | 0.3 | 0.2 | 0.4 | 0.8 | -0.02 | 0.910 | 33 |
| Brown-cheeked Fulvetta | 1.0 | 3.3 | 1.7 | 2.0 | 2.0 | 2.3 | 2.7 | 4.3 | 3.5 | 4.0 | 1.8 | 1.9 | 3.0 | 1.4 | 0.12 | 0.547 | 36 |
| Malabar Whistling Thrush | 0.4 | 0.3 | 0.4 | 0.4 | 0.1 | 0.2 | 0.2 | 0.8 | 0.3 | 0.5 | 0.4 | 1.1 | 0.4 | 0.4 | 0.24 | 0.240 | 39 |
| Crested Serpent Eagle | | | | | | | | | | | | 0.1 | | | 0.26 | 0.192 | 40 |
| Plain Flowerpecker | 0.1 | 0.3 | 0.5 | 0.4 | 0.3 | 0.3 | 1.0 | 0.6 | 0.4 | 0.3 | 0.1 | 0.6 | 0.7 | 0.9 | 0.33 | 0.104 | 42 |
| Little Spiderhunter | 0.1 | 0.2 | 0.2 | 0.2 | | | 0.3 | 0.9 | 0.4 | 0.7 | 0.2 | 0.5 | 0.4 | 0.5 | 0.46 | 0.023 | 49 |
| Indian Scimitar Babbler | 0.3 | 0.1 | 0.3 | 0.4 | 0.9 | 0.6 | 0.6 | 1.2 | 1.1 | 0.9 | 0.8 | 0.4 | 1.1 | 1.1 | 0.49 | 0.014 | 50 |
| Wynaad Laughingthrush* | | 1.4 | | | 5.0 | | | 0.8 | 0.8 | | | | | 1.0 | -0.03 | 0.887 | 32 |
| Common Flameback | | | 0.1 | 0.1 | | 0.1 | | 0.1 | 0.2 | 0.1 | 0.1 | | | | -0.01 | 0.944 | 34 |
| White-bellied Woodpecker | | | | 0.1 | | | | | | 0.1 | | | | | -0.04 | 0.831 | 30 |
| Red Spurfowl (a) | | | | | | | 0.1 | | | | | | | | -0.03 | 0.885 | 31 |
| Dark-fronted Babbler | | 0.1 | 0.5 | | | | 0.2 | | 0.5 | 0.8 | 0.4 | 0.2 | 0.4 | 0.4 | 0.35 | 0.080 | 43 |

| Common name | A | B | C | D | E | F | G | H | I | J | K | L | M | N | Tau | P | Rank |
|---|---|---|---|---|---|---|---|---|---|---|---|---|---|---|---|---|---|
| Emerald Dove | | | 0.1 | | | 0.1 | | 0.2 | 0.2 | 0.1 | | 0.1 | 0.1 | 0.1 | 0.36 | 0.072 | 45 |
| Black-throated Munia | | | 0.2 | | | | | | | | | | | 0.2 | 0.09 | 0.670 | 35 |
| Orange-headed Thrush | | | 0.1 | 0.1 | | 0.1 | | 0.2 | 0.2 | 0.4 | 0.4 | 0.1 | 0.1 | 0.2 | 0.55 | 0.006 | 53 |
| Black Bulbul | | | | | 0.2 | 0.1 | 0.6 | 0.7 | 1.8 | 1.7 | 0.2 | 0.7 | 1.3 | 1.1 | 0.66 | 0.001 | 56 |
| Oriental White-eye | | | | | 0.8 | 1.2 | 3.1 | 6.1 | 5.7 | 5.0 | 6.2 | 11 | 8.9 | 7.6 | 0.83 | 0.000 | 58 |
| Bar-winged Flycatcher-shrike | | | | | | 0.2 | 0.2 | 0.5 | 0.2 | 0.5 | 0.2 | 0.2 | 0.4 | 0.8 | 0.70 | 0.000 | 57 |
| Eurasian Blackbird (b) | | | | | | | | | | 0.1 | | | | 0.1 | 0.15 | 0.469 | 37 |
| Grey-headed Canary Flycatcher | | | | | | | 0.4 | 3.0 | 1.8 | 2.6 | 2.2 | 1.8 | 1.8 | 2.8 | 0.64 | 0.001 | 55 |
| Oriental Dwarf Kingfisher | | | | | | | | | | | 0.1 | | | | 0.20 | 0.311 | 38 |
| Black-lored Tit (c) | | | | | | | | 0.1 | 0.2 | 0.4 | 0.1 | 0.3 | 0.1 | | 0.51 | 0.012 | 51 |
| Nilgiri Flycatcher* | | | | | | | | | 1.2 | 0.8 | 0.4 | | 0.1 | 0.6 | 1.1 | 0.52 | 0.010 | 52 |
| Black-and-Orange Flycatcher* | | | | | | | | | 1.8 | 1.2 | 1.9 | 0.8 | 0.2 | 1.4 | 2.5 | 0.63 | 0.002 | 54 |
| Grey-breasted Laughingthrush* | | | | | | | | | | 0.4 | | | | 2.2 | 0.36 | 0.076 | 44 |
| Speckled Piculet | | | | | | | | | | | 0.1 | | | 0.1 | 0.43 | 0.033 | 48 |
| Red-whiskered Bulbul (d) | | | | | | | | | | | | | 0.3 | | 0.32 | 0.111 | 41 |
| Malayan Night Heron | | | | | | | | | | | | | | 0.1 | 0.38 | 0.060 | 46 |
| White-bellied Shortwing* | | | | | | | | | | | | | | 0.3 | 0.38 | 0.060 | 47 |

*Species endemic to Western Ghats; + species not detected in point counts but occurrence confirmed from opportunistic surveys (or, in rare species, assumed to have been missed). **Incomplete data:** (a) also occurs sporadically in elevations below site G, more infrequently above, (b) also occurs in higher elevation sites, (c) vagrant, seen only once during the study, (d) a non-forest species, occurring at the rainforest edge only in the slightly disturbed site M.